\documentstyle[aps,psfig]{revtex}
\newcommand{\be}{\begin{equation}}
\newcommand{\ee}{\end{equation}}
\newcommand{\bea}{\begin{eqnarray}}
\newcommand{\eea}{\end{eqnarray}}
\newcommand{\cm}{{\cal{M}}}
\newcommand{\mpi}{m_\pi}
\newcommand{\mo}{m_\omega}
\newcommand{\mr}{m_\rho}
\newcommand{\ms}{m_\sigma}
\newcommand{\mph}{m_\phi}
\newcommand{\Go}{\Gamma_\omega}
\newcommand{\Gr}{\Gamma_\rho}
\newcommand{\Gp}{\Gamma_\phi}
\newcommand{\Gs}{\Gamma_\sigma}
\newcommand{\ggo}{g_{\gamma\omega}}
\newcommand{\ggr}{g_{\gamma\rho}}
\newcommand{\ggp}{g_{\gamma\phi}}
\newcommand{\gwrp}{g_{\omega\rho\pi}}
\newcommand{\gprp}{g_{\phi\rho\pi}}
\newcommand{\gpnn}{g_{\pi NN}}
\newcommand{\gsnn}{g_{\sigma NN}}
\newcommand{\gsrr}{g_{\sigma\rho\rho}}
\newcommand{\eps}{\epsilon}
\newcommand{\bu}{\bar u}
\newcommand{\hl}{\overline}
\begin{document}
\draft
\bibliographystyle{unsrt}
\title{Role of in-medium hadrons in photon-nucleus reactions : shadowing and 
dilepton spectrum }
\author{Jan-e Alam$^{a}$, Sanjay K. Ghosh$^{b}$,
Pradip Roy$^{c}$, Sourav Sarkar$^{a}$}   
\address {$^a$ Variable Energy Cyclotron Centre, 1/AF, Bidhannagar,
Kolkata 700 064, INDIA}
\address {$^b$ Department of Physics, Bose Institute, 93/1, A. P. C. Road,
Kolkata 700 064, INDIA}
\address {$^c$ Saha Institute of Nuclear Physics, 1/AF, Bidhannagar,
Kolkata 700 064, INDIA}
\maketitle
\vskip 0.5in
We study the effect of in-medium hadronic properties 
in photon nucleus interactions in the context of shadowing as well
as the dilepton spectrum for incident photon energies in the range 1.1- 3 GeV. 
A reasonable agreement with the experimental data for shadowing is 
obtained in a scenario of downward spectral shift of  the hadrons.  
We show that distinguishable features for in-medium changes of the hadronic 
properties can  be observed experimentally through the dilepton spectrum 
by judicious choice of target nuclei and incident energy of photons.


\section{Introduction}
The study of in-medium properties of hadrons has been a field of great interest 
for quite some time. The recent renewed interest in this area of physics
is mainly due to various results available from
relativistic heavy ion collision  experiments performed at the  Super
Proton Synchrotron at CERN and Relativistic Heavy 
Ion Collider at Brookhaven National Laboratory.
Particularly a series of studies by the CERES/NA45, HELIOS-3 and NA50 
collaborations at SPS with S+Au, S+W, Pb+Au and Pb+Pb 
collisions have largely 
been interpreted as the indication of a downward shift of the \( \rho^{0} \)
mass in the nuclear medium \cite{ceres,helios}. The invariant mass 
spectrum for dilepton production shows a large enhancement at low invariant 
mass regions. These excess dileptons are thought to originate from the decay of 
vector mesons with reduced mass \cite{Li,cassing1} in the medium.

In a contrasting view, some authors explained this enhancement 
as a manifestation of the broadening of the $\rho$ spectral function due 
to its coupling
with baryonic resonances~\cite{rapp1,cassing2,friman1}.
It was initially suggested that the drop in \( \rho \)
mass is mainly governed by the chiral symmetry restoration. But in a theoretical
study, combining chiral SU(3) dynamics with Vector Meson Dominance (VMD), 
it was shown that the chiral restoration does not demand a drastic reduction 
of vector meson mass in the nuclear medium \cite{klingl}. This model also 
predicted a substantial broadening of \( \rho^{0} \) spectral density
with only a marginal mass reduction. 
In ref.~\cite{sh,ann,abhijit} it was shown that 
in a non-chiral model like Walecka model a drastic reduction 
in \( \rho \) meson mass may be realized. 

Presently the heavy ion 
experiments and the corresponding theoretical studies remain
inconclusive mainly due to the fact that the medium effects here are masked 
by the complicated dynamics both in the initial as well as the final state.
Moreover, the huge multiplicities and background makes it difficult to analyze 
the experimental data unambiguously. 

The hadron-nucleus collisions~\cite{ozawa}, 
though seem to be less complicated compared to
A-A collisions, will also suffer from the problem of initial as well as 
final state interactions. On the other hand, these difficulties are largely
overcome with the use of photons which do not have the problem of initial state 
interaction. The high flux photon beam, which has compensated for the low
interaction cross section with nuclear matter, complimented by wide angle
multiparticle spectrometers, have made possible a new generation of 
experiments.

Detection of dileptons in experiments simplifies the problem further since 
in contrast to hadronic, dileptonic decay modes in nuclear matter are 
disturbed by the final state interaction. The small cross section  
for secondary interactions of the outgoing electrons 
with nuclear matter makes the dileptons an ideal probe for studying the 
reactions inside the nucleus.
\par
The dominant feature of photon interaction above \( \pi \) production threshold
with  $ 0.14 < E_{\gamma} $ (GeV) $<0.5$,  \( E_{\gamma} \) being the incident
photon energy in GeV, is the \( \Delta \) resonance production. 
Above this, in the
range $0.5 < E_{\gamma}$ (GeV) $< 1$  , photon nucleus interaction can 
mainly be described by baryon resonance production. The 
$\gamma-A$ cross section per nucleon is suppressed compared to 
the $\gamma-N$ cross section in this energy range.
Beyond $\sim 1$ GeV vector meson production becomes dominant and VMD is 
generally used to describe the interaction in this region.
\par
In the present paper we confine our study in the region 
$1< E_{\gamma}$ (GeV) $< 3$ to understand the role of in medium hadrons
in shadowing as well as in the dilepton spectrum from \( \gamma \)-A 
reaction.  Our endeavour to study shadowing as well as dilepton spectrum is 
important due to following reasons.  
\par
Firstly, the photo-nuclear data at low photon
energies for different nuclei seem to indicate an early onset of shadowing
\cite{bianchi1,bianchi2}. This phenomena has been interpreted as
a signature for lighter \( \rho \) meson in the nuclear 
medium \cite{bianchi3}. In contrast, the authors in ref.~\cite{mosel200} have
claimed that the early onset of shadowing can be understood within simple 
Glauber theory \cite{glauber59,glauber70,yennie71,yennie78} if one takes the
negative real part of the \( \rho \)-N scattering amplitude into account
which corresponds to higher in-medium \( \rho \) mass. In \cite{mosel201},
authors have concluded that the enhancement of shadowing at low energies occur
mostly due to lighter \( \rho \) meson as well as intermediate \( \pi^{0} \)
produced in non-forward scattering. 
\par
But the situation is rather more involved as we have shown in our earlier 
work~\cite{our}. There are several factors which might contribute to the
mismatch of theory and experiment. Glauber model, which was originally
developed for high energy scattering, might be improved by taking care of
the approximations inherent in it. Furthermore, improved estimate of the
parameters of the Glauber model as well as multiple scattering, namely
cross section/scattering amplitudes, might be able to give a better fit
to the experimental data. In Ref.~\cite{our}  we have addressed all these
related problems to show that a good description of photoabsorption data
is not unambiguous and there are some serious constraints in the proper 
understanding of the experimental results. On the other hand, if the 
dropping vector meson mass is the cause of early onset of the shadowing, 
it should be observed in the dilepton spectrum at low energies as well.
In fact, the theoretical analysis of the dilepton spectrum is intimately
related to the shadowing, as the photon-nucleon cross section inside a
nucleus has to be multiplied by the effective number of nucleons {\it i.e.}
$A_{eff}$ ($<A$, being the mass number) 
of the nucleus, as obtained from shadowing studies, to obtain 
the final photon-nucleus cross section. Hence a study of the dilepton spectrum
might put some additional constraints on the shadowing. The dilepton 
study itself may be less ambiguous as a good description of 
photon-nucleon data gives 
{\it a-priori} justification of the models used.
\par
Recently, a large reduction of \( \rho \) mass in the nuclear medium has 
also been
reported by TAGX collaboration \cite{tagx}, inferred from the
dipion spectrum. \( \rho \) meson is certainly the best candidate to study
the medium effects since due to its short life time and decay length a large
portion of the \( \rho^{0} \) mesons produced will decay inside
the nuclear medium. But the detection of dipion is disadvantageous  as
they suffer final state interactions. 
Though $e^+e^-$ branching
ratio is small,
the absence of final state interactions
makes the analysis easier. Of course, there will be a large background
mainly coming from the QED processes like Bethe-Heitler~\cite{bethe},
 but these can be
eliminated with suitable cuts in the final spectrum~\cite{exptcut}.
\par
In the present paper, we have studied, the propagation of the
produced vector meson inside the nucleus. The snap shots at different
times help us to understand the microscopic dynamics in
 \( \gamma \)-A collisions. It
also reveals the fact that \( \omega \) and \( \phi \) mesons 
in contrast to the $\rho$ mostly decay
outside the nuclear environment. The path length traversed by the vector
meson gives a direct correspondence of the dilepton spectrum and the 
medium density at which it has decayed.
\par
The energy range considered here is similar to the proposed
experiment at CEBAF \cite{g7coll}, where incoherent photoproduction of 
vector mesons would be studied, with deuterium, carbon, iron and lead 
as targets. Though, there have been theoretical studies of coherent 
photoproduction earlier, a detailed study 
with in-medium effects was lacking. The present work shows that even 
in coherent processes, the medium effects would show up in an appropriate
kinematic window.  Furthermore, as the Bethe-Heitler
contribution can be suppressed with an energy cut on final 
dileptons, the results presented here will give a fair idea regarding 
the quantitative contribution from 
other coherent processes in the presence of in-medium hadrons.
\par
The paper is organized as follows. In section II, we have 
described the phenomena of shadowing. 
Section III is devoted to the study of the propagation of vector mesons inside
the nucleus and their decay to lepton pairs.
Finally in section IV we present summary and discussions. 

\section{Shadowing in photoabsorption}
The photonuclear cross section has been found to be suppressed compared 
to the photon-nucleon interaction for photon energies, $E_\gamma>1$ GeV.
This phenomenon is attributed to the phenomena of the  nuclear shadowing.  
Here we would like to consider photon  energies above 1 GeV where the 
the dominant process is the production 
of vector mesons in the initial state  and the main tool for theoretical 
description is VMD.

\subsection{Shadowing in \( \gamma \)-A reaction}
If a photon impinges on a nucleus then,  one would naively expect
that the nucleus being transparent to the photon each
nucleon  will have  equal probability to interact with the beam.
In such a case the total cross section for \( \gamma \)-A
reaction would be given by
\begin{eqnarray}
\sigma_{\gamma\,A}=A \sigma_{\gamma N}
\end{eqnarray}
On the other hand, if the photon interacts with nucleus through its hadronic
components, then, just like hadron-nucleus interaction, the photon's 
initial interaction will be principally with nucleons in the incoming
side of the nucleus~\cite{feynman}. 
The nucleons further along the photon trajectory 
do not contribute to the total cross section.  Hence the total 
cross section is expected to be smaller compared to the previous case,
{\it i.e.} \( \sigma(A) < A \sigma_{\gamma -N} \). This phenomena,
observed both in hadron-nucleus as well as \( \gamma \)-nucleus 
interactions, is generally known as shadowing. In case of 
\( \gamma \)-A collision, such a picture would be valid only if the
hadronic components of the photon are long lived. This is quantified 
in terms of the coherent length described later.
The effect of shadowing can be 
written as,
\begin{eqnarray}
\frac{A_{eff}}{A}=\frac{\sigma_{\gamma A}}{A \sigma_{\gamma N}}=
1 + \frac{\delta \sigma_{\gamma A}}{A \sigma_{\gamma N}}
\label{shadow}
\end{eqnarray}
where \( \sigma_{\gamma A} = A \sigma_{\gamma N} 
+ \delta \sigma_{\gamma A} \) consists of the incoherent scattering of the
photon from individual nucleons and a correction due to the 
coherent interaction with several nucleons.
In the present section we will 
discuss the medium effects on shadowing in \( \gamma \)-A reactions 
in the frame work of Glauber formalism \cite{glauber59,glauber70} and multiple 
scattering \cite{mosel201,weis,piller} along with VMD \cite{sakurai}.

\subsubsection{Kinematics and coherence length}
We consider a photon of energy $E_{\gamma L}$ colliding with a 
nucleus at rest.  The nucleons inside the nucleus move with a 
Fermi momenta $p_{F}$ or energy $E_F=\sqrt{p_F^2+m_N^{*2}}$, 
which is a function of space co-ordinate through 
the density $n(r)$. The photon energy in the nucleon rest frame then 
becomes,
\begin{eqnarray}
E_{\gamma} = \gamma_{F} E_{\gamma L} (1-\beta_{F} \cos\theta_{L}),
\label{egamma}
\end{eqnarray}
$\beta_{F}=p_{F}/E_{F}$
and $\theta_{L}$ being the angle between incident photon and the Fermi momenta.
The total invariant energy $s$  can then be written as,
\begin{eqnarray}
s &=& (p_\gamma + p_F)^2 \nonumber \\
&=& {m_{N}^{*}}^2 + 2 \gamma_{F} m_{N}^{*} E_{\gamma L} (1-\beta_{F}\cos\theta_{L}),
\label{ss}
\end{eqnarray}
where $m_N^*$ is the effective nucleon mass inside the nucleus.
The modification of hadronic masses in nuclear environment
have been studied in different models \cite{abhijit,embr,shl,saito}.
In the present study we consider two possible approaches for the
effective mass of the nucleon inside a nucleus: (i) universal
scaling scenario (USS)~\cite{embr} and (ii) Quantum Hadrodynamical Model
(QHD) ~\cite{vol16}.
In case of (i) the hadronic masses ($m_H$), except those of 
pseudo-scalar mesons, 
vary with the nuclear density $n(r)$  as
\begin{equation}
\frac{m_H^*}{m_H}=1-0.2x,
\label{ussm}
\end{equation}
where $x=n(r)/n_0(r)$ and $n_0(r)$ is the normal nuclear matter density
($\sim 0.15$ fm${-3}$).

In the QHD model the effective masses of nucleons and vector
mesons are calculated using standard techniques of thermal field theory
~\cite{ptp,ann}
and parametrized as a function of $n(r)$ as follows:
\begin{eqnarray}
\frac{m_H^*}{m_H}=1+\sum_{j=1}\,a_j\,x^j\,.
\label{qhdm}
\end{eqnarray}
For nucleons  $a_1=-0.351277$ and $a_2=0.0766239$;
in case of $\rho$, $a_1=-1.30966$, $a_2=1.78784$, $a_3=-1.17524$ and
$a_4=0.294456$ and finally for $\omega$, $a_1=-0.470454$, $a_2=0.313825$
and $a_3=-0.0731274$. No medium effects on the $\phi$ meson is considered
as it is expected to be small in QHD~\cite{kuwabara}.
The total invariant energy available for $\gamma N$ scattering now 
depends on the position of the participating nucleon through the 
effective nucleon mass which in turn depends on the density, $n(r)$ or 
the Fermi momentum $p_F$.

The vector meson produced inside the nucleus will have an effective mass
which will depend on the density of the nuclear medium as seen by the 
meson.  A direct effect of these mass changes would be
reflected in the coherence length. The coherence length or the formation
length ($l_c$) is the length scale of the hadronic component of the photon 
inside the nucleus. In other words, this corresponds to the time 
scale of the fluctuation between the bare photon and the hadronic 
component of the physical photon. When $l_c$ is small, the hadron 
mediated interaction may become indistinguishable from bare photon 
interaction and there will not be any shadowing. In the present case, 
$l_c$ is a function of the radial distance inside the nucleus. For 
the vector meson with effective mass $m_{V}^{*}$, the coherence length 
is,
\begin{eqnarray}
l_c = \frac{1}{E_{\gamma} - \sqrt{E_{\gamma}^{2} - {m_{V}^{*}}^2 }}
\,\sim\,\frac{2E_\gamma}{m_V^{*2}},
\label{coherence}
\end{eqnarray}
where $E_{\gamma}$ itself depends on the position of the struck nucleon
through Eq.~(\ref{egamma}) as mentioned before. 

\subsubsection{Multiple Scattering Formalism}
The formalism for the multiple scattering is based on the optical theorem
\cite{weis,gribov}. The nuclear photoabsorption cross section in terms
of amplitudes (\( \cal A \)) of the multiple scattering of the projectile 
with the nucleons inside the nucleus can be written as \cite{piller},
\begin{eqnarray}
\sigma_{\gamma A}=\frac{1}{2m_{N}k} Im \sum^{A}_{n=1}\,{\cal{A}}^{(n)}
\label{multi1}
\end{eqnarray}
where $k$ is the momentum of the photon and $n$ corresponds to the number
of nucleons participating in each multiple scattering process.
The \( n=1 \) term, {\it i.e.}, \( {\cal A}^{(1)} \) is the amplitude of 
forward scattering of a photon with one bound nucleon and corresponds to
the incoherent part in Eq.~(\ref{shadow}). The $n_{\rm th}$ order scattering
amplitude \( {\cal A}^{(n)} \) corresponds to the process where the 
incoming photon produces a hadronic state \( X_{1} \) at the first 
nucleon, \( X_{1} \) propagates freely upto second nucleon where a 
hadronic state \( X_{2} \) is produced which then propagates freely 
to the third nucleon. This process continues till the hadronic state
\( X_{n-1} \) scatters into outgoing photon from $n_{\rm th}$ nucleon. 
The nucleus
is assumed to stay in the ground state which means that there is no
energy transfer to the \( n \) nucleons. Hence the momentum transfer
to the n$_{\rm th}$ nucleon is
\begin{eqnarray}
\vec{q}_{n}=-\sum^{n-1}_{i=1} \vec{q_i}
\end{eqnarray}
The expression for \( {\cal A}^{(n)} \) is,
\begin{eqnarray}
i{\cal A}^{(n)}=\frac{A!}{(A-n)!} \prod_{i=1}^{n-1}
\left[\int \frac{d^3 q_{i}}{2 m_{N} (2\pi)^3} \right]
F(\vec{q_{1}}) \cdots F(\vec{q_{n}}) i{\cal V}^{(n)}(\{ \vec{q_i}
\} )
\end{eqnarray}
where
\begin{eqnarray}
i{\cal V}^{(n)}&=&\sum_{X_i} i{\cal M}_{\gamma X_i}(\vec{q_1})
\frac{i}{\nu^2 - (\vec{k} - \vec{q_1})^2 - m^2_{X_1} - 
\Pi_{X_1}(\nu^2 - (\vec{k} - \vec{q_1})^2)} 
i{\cal M}_{X_1 X_2}(\vec{q_2}) \nonumber \\
&\times& \frac{i}{\nu^2 - (\vec{k} - (\vec{q_1}+\vec{q_2}))^2 -
m^2_{X_2} - \Pi_{X_2}(\nu^2 - (\vec{k} - (\vec{q_1} + \vec{q_2}))^2)}
\nonumber \\
&\times& \cdots i{\cal M}_{X_{n-1} \gamma}(\vec{q_n})
\end{eqnarray}
\( m_{X_i} \) and \( \Pi_{X_i} \) denote the mass and the vacuum self
energy of the intermediate hadronic state \( X_i \). 
\( {\cal M}_{a b} \) corresponds to the invariant amplitude
for the process \( a + n \rightarrow b+ n \) and \( F(\vec{q_i}) \)
is the nuclear form factor,
\begin{eqnarray}
F(\vec{q_i})=\frac{1}{A} \int {d^3}x e^{i\vec{q_i}.\vec{x}}
n(\vec{x})
\end{eqnarray}
where \( n(\vec{x}) \) denotes the nucleon number density.
Under eikonal approximation and neglect of the width
of the vector meson in the intermediate state the above formalism
reduces to the Glauber's formula. In the present context, in order to
understand the role of in-medium hadrons, we modify the Glauber's formula. 
The photon entering the nucleus at an impact parameter $b$ produces a vector 
meson at position $z_{1}$.   
Inside the nucleus, the coherence length $l_c$, in general, would
be different at $z_1$ ($l_{c1}$) and $z_2$ ($l_{c2}$) as the 
different densities will yield different masses. 
The expression for the shadowing part of the cross section 
is then given by~\cite{yennie71},
\begin{eqnarray}
&&\delta \sigma_{VA}=\frac{g_V^2}{4\pi\alpha}\delta\sigma_{\gamma\,A} \nonumber\\
&=& \frac{1}{2k k_{V}}\int{d^2b} \int dz_{1} \int dz_{2}
\exp \left[- \frac{1}{2} \int \sigma_{VN}(z') n(b,z') dz' \right] \nonumber \\
& \times &
k_{V}(z_{1}) \sigma_{VN}(z_{1}) k_{V}(z_{2}) \sigma_{VN}(z_{2}) n^{(2)}(b,z_{1},z_{2})
\nonumber \\
& \times & \left[ \left( \alpha_{V}(z_{1}) \alpha_{V}(z_{2}) - 1 \right)~~ \cos\left(
(\frac{z_{1}}{l_{c1}} -
\frac{z_{2}}{l_{c2}}) \right. \right. \nonumber \\
&+& \left. \frac{1}{2} \int_{z_1}^{z_2} \alpha_{V}(z') \sigma_{VN}(z') n(b,z') dz' \right)
- \left( \alpha_{V}(z_{1}) + \alpha_{V}(z_{2}) \right)  \nonumber \\
& \times & \left. \sin\left( (\frac{z_{1}}{l_{c1}} - 
\frac{z_{2}}{l_{c2}}) + \frac{1}{2} \int_{z_1}^{z_2} \alpha_{V}(z') \sigma_{VN}(z')
n(b,z') dz'\right) \right],
\label{deltasig}
\end{eqnarray}
where $\alpha_{V}={Re f_{VV}}/{Im f_{VV}}$ is the ratio of the real to the
imaginary part of the $VN$ forward scattering amplitude~\cite{yennie78}. 
$\sigma_{VN}$ 
is the $V-N$ scattering cross section~\cite{yennie78} and $k_{V}$
is the wave vector of the vector meson.
The attenuation of the vector meson amplitude is described by  the 
exponential factor.
We have included 2-body correlation in the 
two-particle density as~\cite{mosel200},
$n^{(2)}(b,z_{1},z_{2}) = n(b,z_{1}) n(b,z_{2})
[1-j_{0}(q_{c} |z_{1}-z_{2}|)]$, 
where $q_{c}=780$ MeV and $j_{0}$ is the spherical Bessel function. 
The theoretical results of our calculations have been compared with 
experimental data. The extraction of experimental numbers has been 
discussed in ref.~\cite{our}. 
\par
In the present paper we have discussed the results of shadowing for
Carbon (C) and lead (Pb) nuclei. Depending on the size of the nucleus
we have used two different density distributions; for $A<16$ 
the shell model density profile of Ref.~\cite{shaw}  
and for heavier nuclei ($A>16$) the density profile from 
ref.~\cite{eskola} has been used.
According to eq.~(\ref{egamma}), $E_\gamma$ is a function
of angle, $\theta_L$ for non-zero $p_F$.
The results which are presented here have been averaged 
over all possible values of $\theta_L$.  
We have found that the effect of Fermi momentum 
in the kinematics (through eq.~(\ref{egamma})) is small.

\begin{figure}
\centerline{\psfig{figure=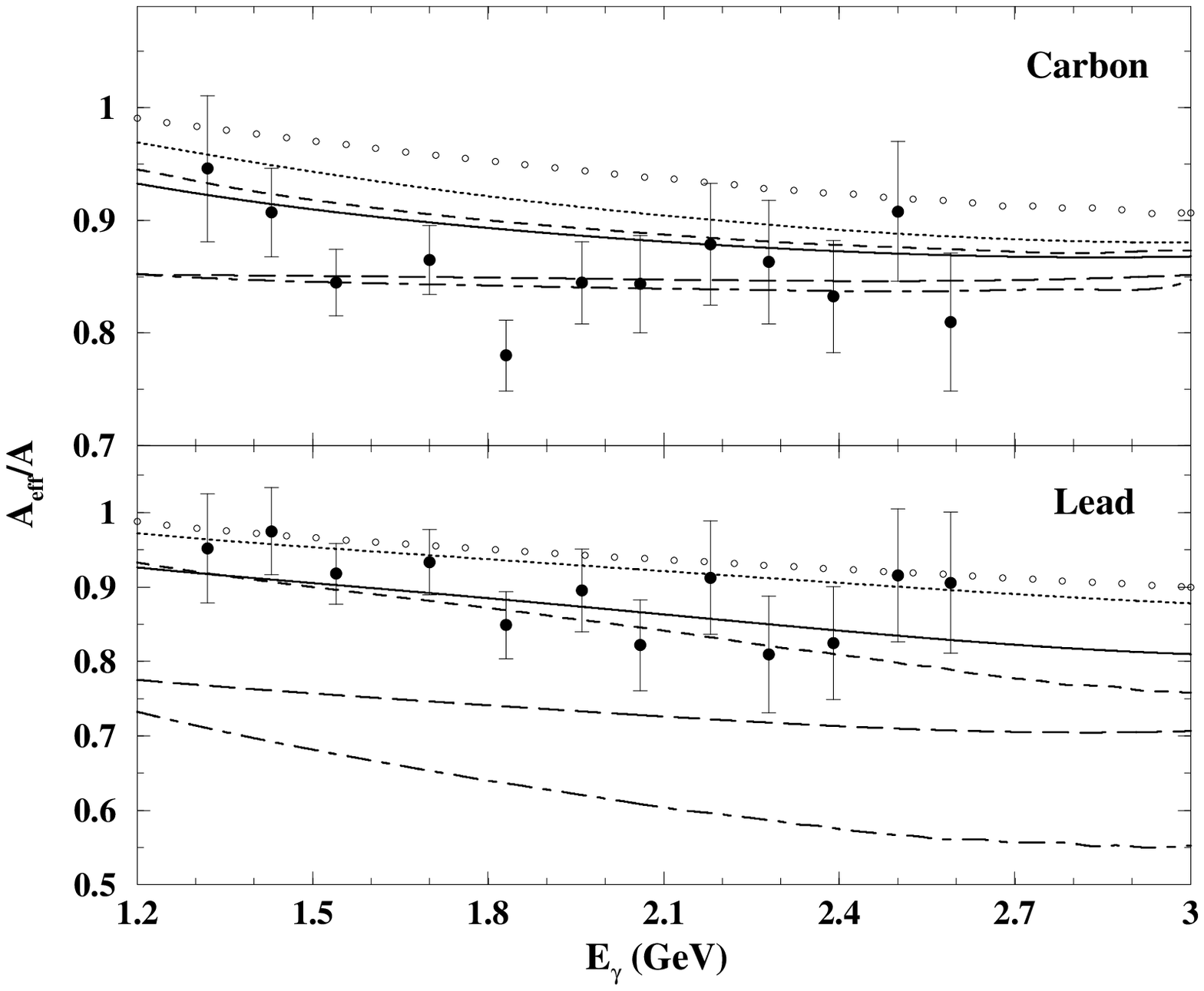,width=9cm,height=12cm}}
\caption{$A_{eff}/A$ for C and Pb nuclei as a function of photon
energy.  We show the results for both 
multiple scattering approach and Glauber model. The dotted, long-dashed and
solid lines indicate calculations using Glauber model for vacuum, QHD and
USS respectively. The circles, dot-dashed  and
short-dashed lines correspond to the same in the multiple scattering approach.}
\label{fig1}
\end{figure}

Fig.~(\ref{fig1}) shows the variation of $A_{eff}/A$ with $E_{\gamma}$ 
for C and Pb nuclei. We have used Glauber model and
multiple scattering approach for the energy ranges 
considered here. We find that the
USS gives a better description of the data both in the
multiple scattering approach  and Glauber model than the scenario
with vacuum properties of hadrons.
In QHD the drop of hadronic masses
being larger, in general, the data is underestimated. On an average, 
our results show that the experimental data over the entire range of 
photon energy under consideration are reasonably well reproduced by 
the downward shift of the spectral function within the USS.

The results shown here have been found to be sensitive to the pole mass and 
largely insensitive to the broadening of the spectral function. But, more 
importantly, the results depends crucially on model (Glauber or 
multiple scattering) parameters, and less on the approximations 
({\it e.g.} eikonal approximation) involved.
Since the USS describe the data 
on shadowing reasonably well for the energy range $1<E_\gamma$ (GeV)$<3$, we
will evaluate the dileptons within this energy range in the USS.
For comparison we will give results for both vacuum as well as QHD scenario.

\section{Dileptons from \( \gamma \)-A collision}
The photoproduction of vector mesons, specifically \( \rho \) and 
\( \omega \) mesons, have been studied before \cite{leith,friman2}.
Recently dilepton spectrum have also been studied in photon-nucleus 
interaction using Boltzmann-Uheling-Uhlenbeck (BUU) formalism
\cite{effen}. 
In this work we consider 
$t$ channel processes (above resonance region) to study the 
effects of in-medium hadrons on the dilepton spectrum. 
The $t$ channel diagrams considered in the present work
provides 
a reasonable description of vector meson photoproduction data 
as we will show below (see also \cite{friman2}). It is well known
that to evaluate the lepton pair production from the leptonic decay of the 
vector meson in a nuclear medium all the processes which can either
create or annihilate the vector meson under consideration has to 
be included in the imaginary part of its self energy 
(which is proportional to the in-medium width).
   
\subsection{Coupling constants and form factors for t-channel processes}
\begin{figure}
\centerline{\psfig{figure=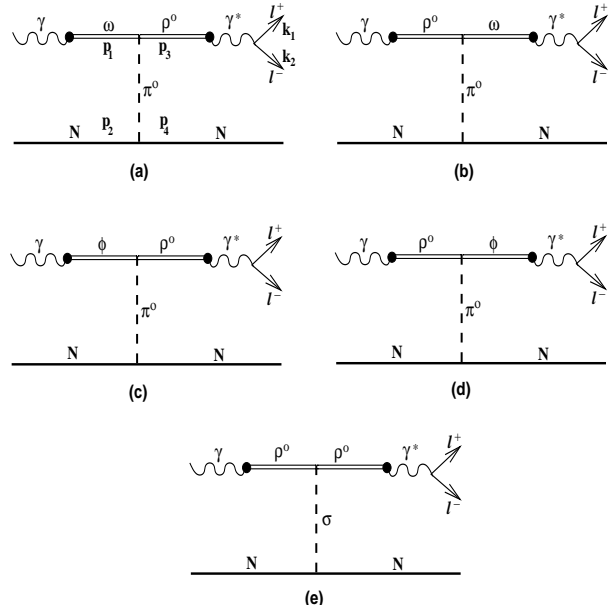,width=8cm,height=8cm}}
\caption{Feynman diagrams for t-channel processes}
\label{fig2}
\end{figure}

Here we will describe the different processes and the evaluation of the
required coupling constants. The reactions considered in this paper are
shown in Fig.~(\ref{fig2}). The iso-scalar and iso-vector coupling of
electromagnetic current and vector fields can be written following the 
current field identity \cite{kroll}:
\begin{eqnarray}
{\cal {J}}_{\mu}^{(em)}(I=0) = \frac{e m_\omega^{2}}{g_\omega} \omega_\mu
\nonumber \\
{\cal {J}}_{\mu}^{(em)}(I=0) = \frac{e m_\phi^{2}}{g_\phi} \phi_\mu \nonumber \\
{\cal {J}}_{\mu}^{(em)}(I=1) = \frac{e m_\rho^{2}}{g_\rho} \rho_\mu
\label{current}
\end{eqnarray}
where \( m_\omega \), \( m_\phi \) and \( m_\rho \) are the masses of
\( \omega \), \( \phi \) and \( \rho \) mesons respectively. The couplings
\( g_\omega \), \( g_\phi \) and \( g_\rho \) are evaluated from the
\( e^{+} e^{-} \) partial decay widths of the corresponding mesons and found
to be 5.03, 13.25 and 17.05 respectively.
\par
Processes (a)-(d) go through \( \pi \) and 
process (e) occurs via \( \sigma \) meson exchange. 
The exchange of vector mesons
are not allowed because of the violation of charge conjugation invariance.
The \( \eta \) exchange is not considered as it will be suppressed due to
its larger mass and smaller value of coupling compared to pions.  
For the case of \( \omega \) in the final channel, the
neglect of \( \eta \) is further justified as \( \omega \rightarrow \eta
\gamma \) is two orders of magnitude lower than the \( \omega \rightarrow
\pi^{0} \gamma \). Due to opposite parity
of \( \pi \) and \( \sigma \), diagrams (a)-(d), which interfere among 
themselves, do not interfere with (e)~\cite{friman2}. 
The relative importance of \( \pi \) and 
\( \sigma \) exchange diagrams can be assessed from the radiative 
transition of vector mesons. The branching ratio of 
\( \omega \rightarrow \pi^0 \gamma \) is about 
an order of magnitude  larger than 
than that of \( \omega \rightarrow \pi^+ \pi^- \gamma \) 
which has an upper limit
3.6 \( \times 10^{-3} \)~\cite{pdg} 
which implies a dominance of \( \pi \) exchange
for \( \omega \) in the final channel. In contrast, for \( \rho \)
meson \( \rho \rightarrow \pi^0 \gamma \) is almost one order 
of magnitude lower than
\( \rho \rightarrow \pi^+ \pi^- \gamma \) showing the importance
of \( \sigma \) exchange over \( \pi \) for reactions involving $\rho$
in the final channel. 
\par
Let us now consider the different interactions needed to evaluate the diagrams. The \( \pi - N \) and \( \sigma - N \) interactions are
\begin{eqnarray}
{\cal {L}}_{\pi NN} = -i g_{\pi NN} \bar{N} \gamma_{5} 
(\vec{\tau}.\vec{\pi})N   \nonumber \\
{\cal {L}}_{\sigma NN} = g_{\sigma NN} \bar{N} N \sigma 
\label{pisign}
\end{eqnarray}
where \( g_{\pi NN} = 13.26 \) and \( g_{\sigma NN} = 10.03 \) 
\cite{friman2}. The corresponding vertex form factors are
\begin{eqnarray}
F_{\pi NN} = \frac{\Lambda_{\pi(\sigma)}^2 - m_{\pi(\sigma)}^2}{
\Lambda_{\pi(\sigma)}^2 - q^2}
\label{form}
\end{eqnarray}

\begin{figure}
\centerline{\psfig{figure=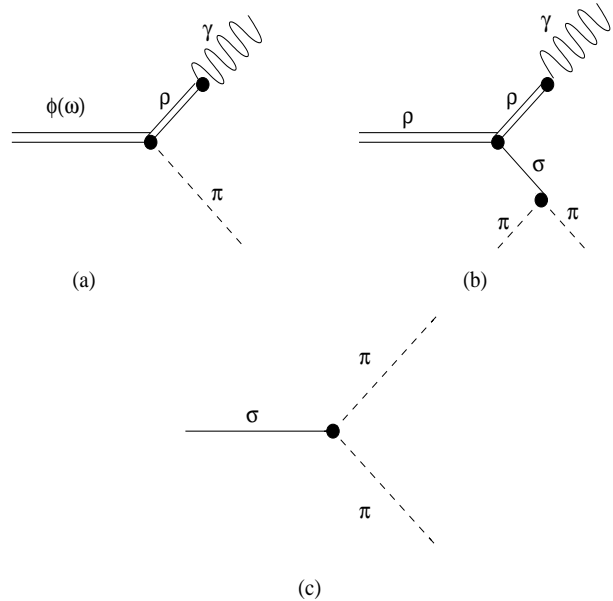,height=8cm,width=8cm}}
\caption{Feynman digram for coupling constants evaluation}
\label{fig3}
\end{figure}
The anomalous $\omega\rho \pi$ interaction~\cite{gsw} is given by
\be
{\cal L}_{\omega \rho \pi} = \frac{g_{\omega \rho \pi}}{m_{\pi}}\,
\epsilon^{\mu \nu \alpha \beta}\,
\partial_{\mu}\omega_{\nu}\,\partial_{\alpha}\rho_{\beta}\,\pi.
\ee

\begin{figure}
\centerline{\psfig{figure=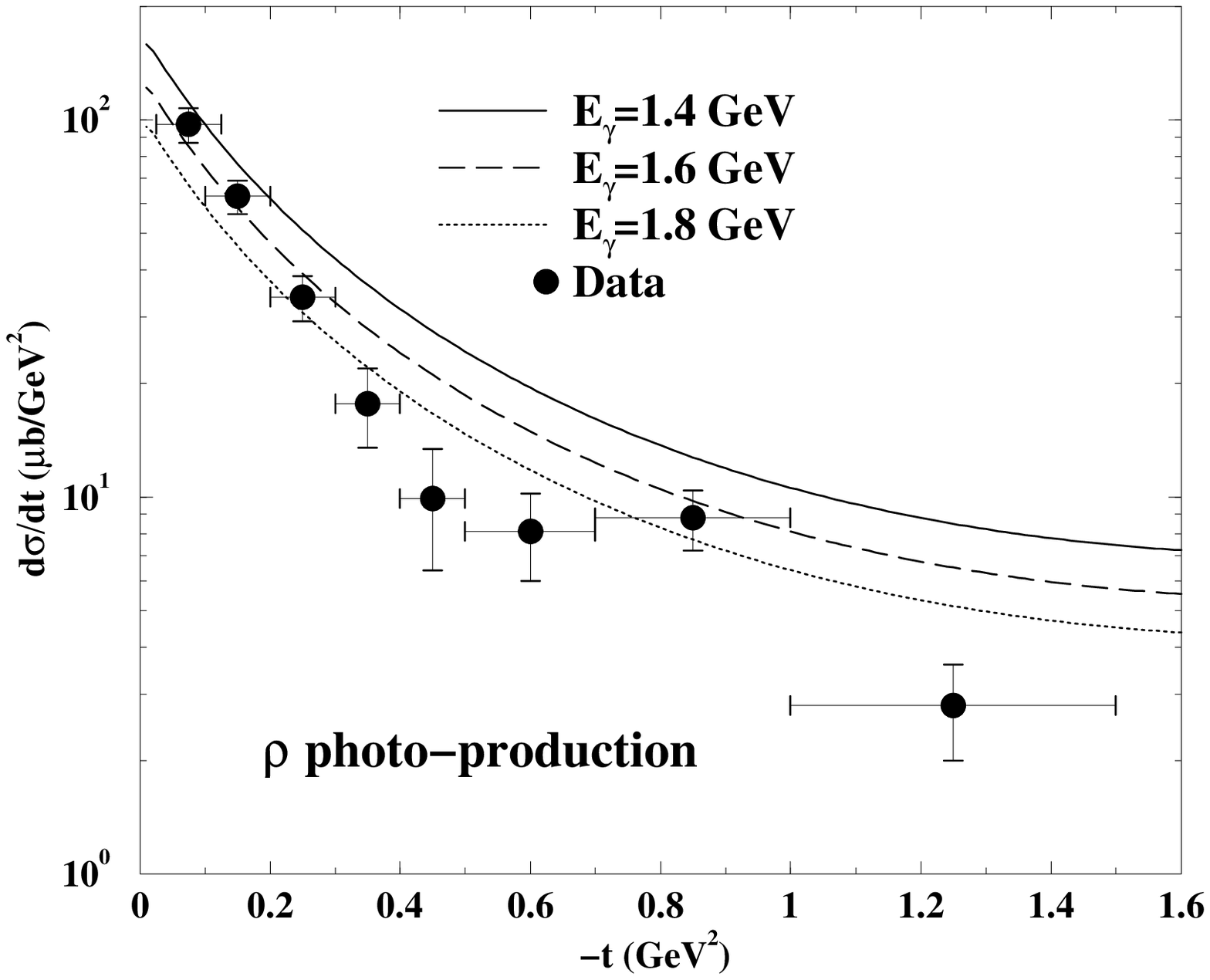,width=8cm,height=8cm}}
\end{figure}
\begin{figure}
\centerline{\psfig{figure=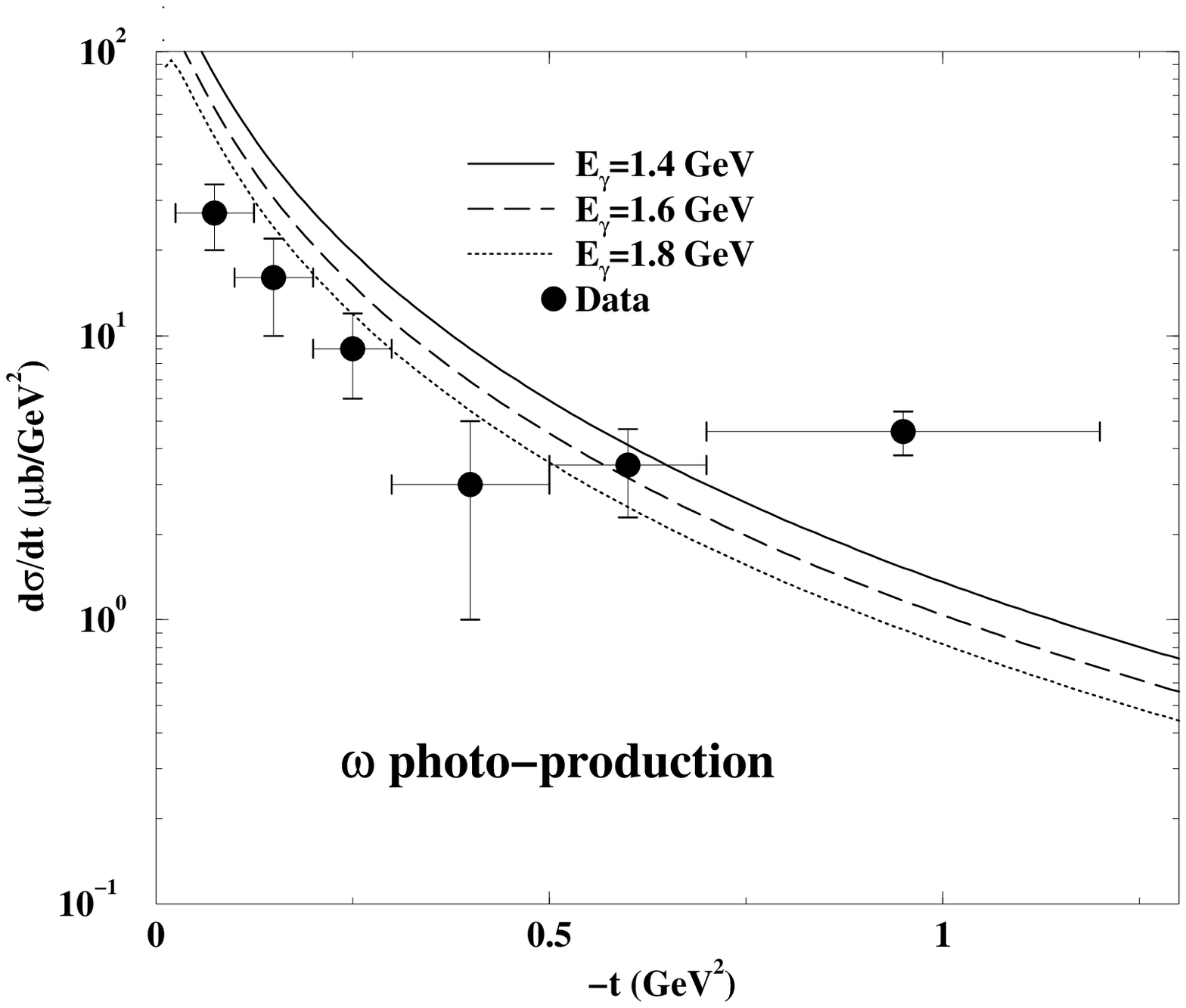,width=8cm,height=8cm}}
\caption{$\rho$ and $\omega$ photo-production cross-sections. The existing
data is compared with the theoretical calculations 
(see{\protect\cite{friman2}} also).
}
\label{fig4}
\end{figure}

The coupling constant $g_{\omega \rho \pi}$ is determined from the decay
$\omega\,\rightarrow\,\pi^0\,\gamma$ (see Fig.~(\ref{fig3}a)) assuming a $\rho$ dominance. The decay width
is given by
\be
\Gamma_{\omega \rightarrow \pi^0 \gamma} = \frac{\alpha\,g_{\omega \rho \pi}^2}
{24 g_{\rho}^2}\,\frac{m_{\omega}^3}{m_{\pi}^2}\,\left(1-\frac{m_{\pi}^2}
{m_{\omega}^2}\right)^3
\ee
We obtain $g_{\omega \rho \pi}$ to be 1.92 using $\Gamma_{\omega \rightarrow \pi^0 
\gamma}$ = 715 keV. The $\phi \rho \pi$ interaction can be written in a 
similar fashion. By using the radiative decay of $\phi$, $g_{\phi \rho \pi}$
comes out to be 0.097.

The $\sigma$-exchange interaction for $\rho$ photoproduction is given by
\be
{\cal L}_{\sigma \rho \rho} = \frac{g_{\sigma \rho \rho}}{m_{\pi}}\,
\partial^{\alpha} \rho^{\beta}\,(\partial_{\alpha} \rho_{\beta}-
\partial_{\beta} \rho_{\alpha})\,\sigma
\label{sigroro}
\ee
The coupling constant $g_{\sigma \rho \rho}=4.6$, is determined from 
$\rho\,\rightarrow\,\pi^+\,\pi^-\,\gamma$ decay width given by 
(see Fig.(\ref{fig3}b))
\be
\Gamma_{\rho \rightarrow \pi^+ \pi^- \gamma} = 
\frac{\pi^2}{2m_{\rho}(2\pi)^5}\,\int\,dE_1\,dE_2\,\langle |{\cal M}|^2
\rangle
\ee
where $E_1$ and $E_2$ are the pion energies.  The spin averaged matrix element 
squared is given by
\be
\langle |{\cal M}|^2\rangle = \frac{8\pi \alpha}{3}\,
\frac{g_{\sigma \rho \rho}^2 g_{\sigma \pi \pi}^2}{g_{\rho}^2}\,m_{\pi}^2
\,\frac{(m_{\rho}-E_1-E_2)^2}{[2m_{\rho}(E_1+E_2)-m_{\rho}^2-m_{\sigma}^2]^2}
.
\ee
The coupling constant $g_{\sigma \pi \pi}$ is obtained as
(see Fig.~(\ref{fig3}c))
\be
g_{\sigma \pi \pi} = \frac{32\pi}{3}\,\frac{m_{\sigma} \Gamma_{\sigma \rightarrow
\pi \pi}}{m_{\pi}^2} (1-4m_{\pi}^2/m_{\sigma}^2)^{1/2} = 17.6
\ee
with $m_{\sigma} = 500$ MeV and $\Gamma_{\sigma \rightarrow \pi \pi} = 300$ MeV.

The \( \sigma \rho \rho \) form factor in monopole form is given by,
\begin{eqnarray}
F_{\sigma \rho \rho}=\frac{\Lambda_{\sigma \rho \rho}^2 - 
m_{\sigma}^2}{\Lambda_{\sigma \rho \rho}^2 - q^2}
\end{eqnarray}

The results for the differential cross sections of \( \rho \) and \( \omega \) 
photoproduction in \( \gamma \)-nucleon collisions along with the experimental 
data for \( \gamma \) energy 1.4-1.8 GeV are shown in Fig.~(\ref{fig4}). 
The cut-off \( \Lambda_{\pi (\sigma)} \) characterizing the 
\( \pi (\sigma) \)-nucleon vertex is 0.7 GeV (1 GeV)~\cite{friman2}.
\( \Lambda_{\sigma \rho \rho} \) = 0.9 GeV is taken from \cite{friman2}.

\subsection{Propagation of vector mesons inside nucleus}

The vector mesons produced inside the nucleus take a finite time to travel
through the nucleus. The modification in their width affects the
probability of these mesons to decay inside the nucleus. To have a
better understanding of this phenomenon, we have studied the propagation
of vector mesons with time after they are produced. 

The minimum photon energy required to produce a vector meson
of mass $M$ can be calculated as follows. The magnitude of the three 
momentum $\mid \vec{p_3} \mid$ of the vector meson produced in a $\gamma$-
nucleon collision is,
\be
\mid \vec{p_3} \mid=\frac{1}{2\sqrt{s}}\lambda^{1/2}(M^2,m_N^2,s)
\ee
where $\lambda$ is the triangular function defined in the appendix A.
Substituting $s=m_N^2+2m_NE_\gamma$ in the above equation we
obtain the threshold for the incident beam energy of the
photon as $E_0=M^2/2m_N+M$. For the production of 
$\rho$ meson of mass 770 MeV $E_0=1.1$ GeV. It should be mentioned
here that if the mass of the $\rho$ reduces in the medium
then this threshold will also reduce.
For $E_0=1.1$ GeV
the $\rho$ mesons will be produced but not $\omega$ and 
$\phi$. Therefore, any interference effects between $\rho$
and $\omega$ will be absent. Moreover, for this incident 
photon energy the $\rho$ will be created almost at rest
so that it decays inside the nucleus. This will lead 
to a very clear signal for the shift of the $\rho$
spectral function in the medium.

Now we will describe the
methodology used to study  vector meson propagation inside the 
medium.  The matrix element
for \( \gamma \)-N collision inside a nucleus would in general depend on
the position $r$ of the nucleon inside the nucleus. More specifically,
the mass and width of the vector meson are
functions of $r$ inside the nucleus and is measured with respect to the center
of the nucleus. The vector meson produced inside the nucleus at
\( \vec{r} \), would propagate inside the nucleus with a 
velocity \( \vec{v} \) (say) and would decay after traveling a distance 
given by \( \vec{r'} = \vec{r} + \vec{v} t \) in time $t$. Now, the 
distance traveled by the vector meson would depend on its total
decay width in the medium,
\begin{eqnarray}
\Gamma_{\rm tot} = \Gamma_0 + \Gamma_{\rm coll},
\label{vmwidth}
\end{eqnarray}
where \( \Gamma_0 \) is the decay width in vacuum and  \( \Gamma_{\rm coll} \)
is the width due to the interaction of the vector meson in the medium.
The increase in the width of the vector mesons reduce their life
time in the medium and hence may allow them to decay within the nucleus. 
The time available to this
vector meson to propagate, before it decays into dileptons, would be 
inversely proportional to \( \Gamma_{\rm tot} \). So one can now write
the distance traveled \( r_l=v t \) as,
\begin{eqnarray}
r_l = \frac{\gamma\,v}{\Gamma_{\rm tot}};~~~~~\gamma=\frac{1}{\sqrt{1-v^2}} 
\end{eqnarray}
where the expression of $v$ is,
\begin{eqnarray}
v=\frac{\mid\vec{p_3}\mid}{E_3}=
\sqrt{1 - \frac{4 m_N^{*2}(r) M^2}{(s + t - m_N^{*2}(r))^2}}
\label{velocity}
\end{eqnarray}
where all the quantities in Eq.~(\ref{velocity}) are evaluated at the 
density of the production point $r$. The point at which the final vector 
meson decays can now be defined as,
\begin{eqnarray}
r' = \sqrt{r^2 + r_l^2 - 2 r r_l \cos{\theta}}
\end{eqnarray}
\( \theta \) being the angle between the velocity vector \( \vec{v} \)
and \( \vec{r} \). 

\par
The vector meson passing through the nucleus may undergo multiple 
collisions with the other particles present in the medium which 
leads to a broadening of its width, which can be estimated in following 
manner \cite{kondratyuk}. Assuming that the vector meson is narrow
and medium density 
is rather low, the intensity of the vector mesons passing through
the vacuum can be described as,
\begin{eqnarray}
|\psi(t_{life})|^2 = |\psi(0)|^2 e^{-\frac{\Gamma_0 t_{life}}{\gamma}}
\end{eqnarray}
In the nuclear medium 
this intensity decreases further due to scattering and absorption 
processes.  In the medium $\Gamma_0$ is replaced by $\Gamma_{tot}$ which
is given by Eq.~(\ref{vmwidth}).
So the net reduction would go as \( \exp(-\Gamma_{\rm tot} t_{life}/\gamma)
\). The increase in width due to collisional broadening in the medium 
is given by,
\begin{eqnarray}
\Gamma_{\rm coll} = n(r) \sigma^* v \gamma
\end{eqnarray}
where \( \sigma^* \) is the total vector meson -
nucleon cross section and \( n(r) \) is the density of the nuclear 
medium.  
The average number of collisions the vector meson suffers in the 
medium can be estimated  to be 3 for Pb
and 1.6 for C at \( E_{\gamma} = 1.1 \) GeV. For \( E_{\gamma} = 2 \)
GeV these numbers are 3 and 1.6 respectively. 
These clearly show that number of
collisions is not very large, even for Pb. 
In that case, the experimental picture, perhaps 
will be closer to the no broadening scenario.
Mass modifications in the medium appears to be the major factor in this 
case.

\subsection{Dilepton spectrum}

The formalism discussed in previous sections have been used to evaluate
\( \gamma \)-A cross section for different scenarios. Here we would
like to point out that at low energies some of the baryonic resonances
get coupled to the \( \rho \) meson. 
It has been shown in ~\cite{kondra} that these couplings make
the $\rho$ spectral function very broad. In the present work
this effect is taken into account through the width of the 
spectral function. To avoid double counting one should 
not separately add dilepton production from these resonances as
these are already included in the spectral function~\cite{weldon}.

Now the matrix element for dilepton production (Fig.~\ref{fig2}) 
is  recasted so as to incorporate the  
co-ordinate  dependence for production and decay points. For a reaction
of the type $p_1 + p_2 \rightarrow p_3 + k_1 + k_2$ this is
generically given by,
\begin{eqnarray}
|{\cal M}|^2 = {\cal A}_{ex} \frac{(k_1 . k_2) \left[ (p_1 . k_1)^2 
+ (p_1 . k_2)^2 \right] (p_2 . p_4 + m_N^2(r))}{\left[ m_{p_1}^4(r) + 
m_{p_1}^2(r) \Gamma_{p_1}^2(r) \right] \left[ (M^2 - m_{p_3}^2(r'))^2 
+ m_{p_3}^2(r') \Gamma_{p_3}^2(r') \right] \left[ (t-m_{ex}^2(r))^2 +
\Gamma_{ex}^2(r) m_{ex}^2(r) \right]}
\label{matrixrp}
\end{eqnarray}
where \( m_{ex} \) and \( \Gamma_{ex} \) are the mass and decay width of 
the
exchanged particles ( \( \pi \) and \( \sigma \) in the present
case). The masses of the initial state vector mesons 
( \( m_{p_1} \)) produced via VMD and the nucleon mass
\( m_N \) depend on $r$ whereas mass \( m_{p_3} \) and width
\( \Gamma_{p_3} \), for the final vector mesons are evaluated 
at $r'$. ${\cal A}_{ex}$ 
contains all the factors coming from 
the different couplings (see appendix B). 
The effective masses are evaluated using Eq.~(\ref{ussm}) for
universal scaling and Eq.~(\ref{qhdm}) for QHD scenarios.
The 4-momenta \( p_1, p_2, p_3, \) and \( p_4 \) correspond to the initial 
photon, initial nucleon, intermediate vector meson and final nucleon 
respectively. The vector meson with momentum \( p_3 \) finally
decays to lepton pairs of momenta \( k_1 \) and \( k_2 \) via a 
virtual photon.
The differential cross section for such a process can be written as 
(see Appendix for details),
\be
\frac{d\sigma_{l^+ l^-}}{dM^2}=\frac{1}{256\pi^3\lambda(s,0,m_N^2)}
\int_{t_{1min}}^{t_{1max}}\,\frac{dt_1}{\lambda^{1/2}(M^2,0,t_1)}
\,\int_{t_{2min}}
^{t_{2max}}\,dt_2
|\hl{\cm}(M^2,t_1,t_2)|^2
\ee
The detailed forms of the matrix elements \( {\cal M} \) are given 
in the appendix along with 
the expressions as well as allowed regions of $s$, $t_1$ and $t_2$. 
In oder to incorporate medium effects on the dilepton cross-section from
$\gamma-A$ collisions we convolute the above equation with the density profile
of the nucleus concerned. As mentioned earlier, there is shadowing phenomenon
in $\gamma-A$ reactions which affects the dilepton production. Taking these 
facts into account we obtain dilepton production cross-section per nucleon
from $\gamma-A$ reactions as 
\be
\frac{d\sigma^{A}_{l^+ l^-}}{dM^2}=\frac{1}{256\pi^3
\lambda(s,0,m_N^2)}
\,\frac{\int\,d^3r\,\rho(r)\,\int_{t_{1min}}^{t_{1max}}\,\frac{dt_1}{\lambda^{1/2}(M^2,0,t_1)}
\,\int_{t_{2min}}
^{t_{2max}}\,dt_2
|\hl{\cm}(M^2,t_1,t_2)|^2}
{\int\,d^3r\,\rho(r)}
\ee

The invariant mass distribution of the lepton pairs resulting 
from the decay of the vector meson {\it inside} the nucleus is given
by

\be
\frac{d\sigma^{in}_{l^+l^-}}{dM^2}=
\frac{\int\,(1-P(R_A))\,\frac{d\sigma_{l^+l^-}}{dM^2}\,\rho (r)\,d^3r}
{\int \rho (r)d^3r}
\label{decinside}
\ee
where $P(R_A)$ is given by,
\be 
P(R_A)=e^{-MR_A\Gamma_{tot}/\mid\vec{p_3}\mid}
\label{probab}
\ee
is the probability of the vector mesons of mass $M$ 
to decay outside the nuclear
radius $R_A$. In a similar way one can evaluate the dilepton emission
rate {\it outside} the nucleus by replacing $(1-P(R_A))$ by $P(R_A)$ in 
Eq.~(\ref{decinside}).

\begin{figure}
\centerline{\psfig{figure=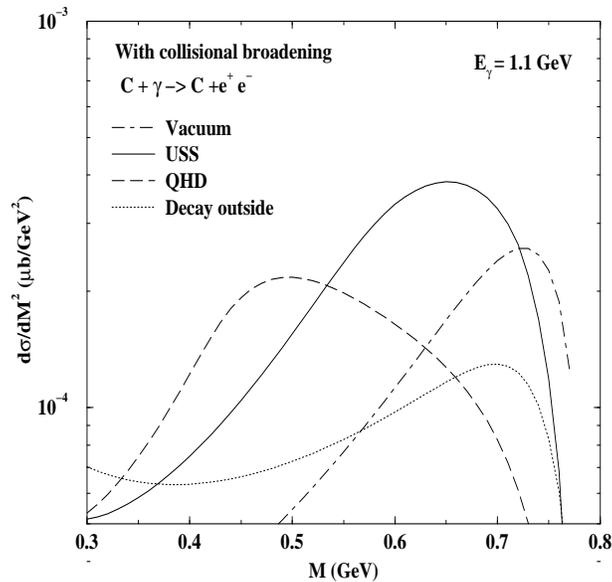,width=8cm,height=8cm}}
\caption{Invariant mass distribution of lepton pairs from $\gamma$-C
collisions at  $E_\gamma=1.1$ GeV. The result indicated by vacuum
corresponds to the mass of the vector meson in vacuum and the 
in-medium width evaluated by using eq.~\protect{\ref{vmwidth}}. 
The curves denoted by USS and QHD correspond to the 
medium dependent masses given by eqs.~\protect{\ref{ussm}}
and ~\protect{\ref{qhdm}} respectively.
}
\label{fig5}
\end{figure}
\begin{figure}
\centerline{\psfig{figure=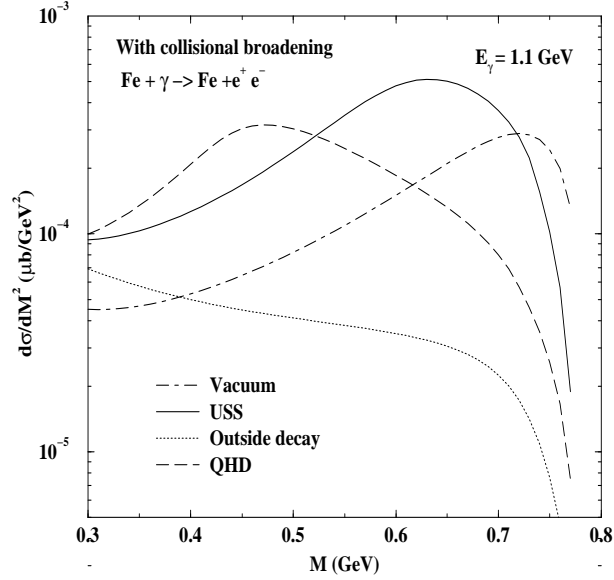,width=8cm,height=8cm}}
\caption{Same as \protect{Fig.\ref{fig5}} 
for  $\gamma$-Fe collisions.
}
\label{fig6}
\end{figure}
\begin{figure}
\centerline{\psfig{figure=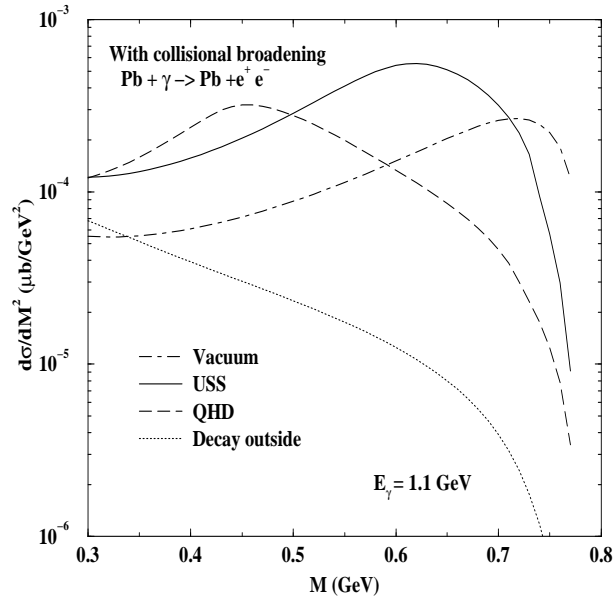,width=8cm,height=8cm}}
\caption{Same as \protect{Fig.\ref{fig5}} 
for  $\gamma$-Pb collisions.
}
\label{fig7}
\end{figure}

In Figs.~\ref{fig5}, \ref{fig6},\ref{fig7} we have depicted the
the invariant mass distribution of lepton pairs for incident
photon energy of 1.1 GeV for C, Fe and Pb nuclei respectively. 
\( d\sigma/dM^2 \)  indicates the  expected dilepton spectrum 
if the vector mesons decay inside or outside depending on the
energy as well as their decay widths. If the meson decays outside,
a sharp peak is to be observed at the \( \rho \) mass of 770 MeV. 
In the case of mesons decaying inside,
one observes distinguishable differences between the vacuum and
the in-medium scenarios.  At $E_\gamma=1.1$ GeV only the $\rho$
meson can be produced with negligible velocity. 
In this case
the probability for the decay of $\rho$ inside the nucleus
is maximum. Hence it will bring the information of the 
in-medium spectral function of the $\rho$ most effectively.
In the present case the width of the $\rho$ due its
interaction with baryonic resonances is given by Eq.~(\ref{vmwidth}) and
the pole of the  spectral function is shifted according to Eq.~(\ref{ussm})
or (\ref{qhdm}). For all the targets (C, Fe and Pb) 
considered here,  the decay probability of the 
$\rho$ meson outside the nucleus is much smaller than its decay
inside. For incident photon energy close to the threshold
the heavier the target less is the decay probability outside 
the nucleus. The effects of the in-medium modification  of the
$\rho$ is clearly visible through dilepton spectra for 
the scenarios, {\it i.e.} USS and QHD 
for the entire range of $M$ considered here.

\begin{figure}
\centerline{\psfig{figure=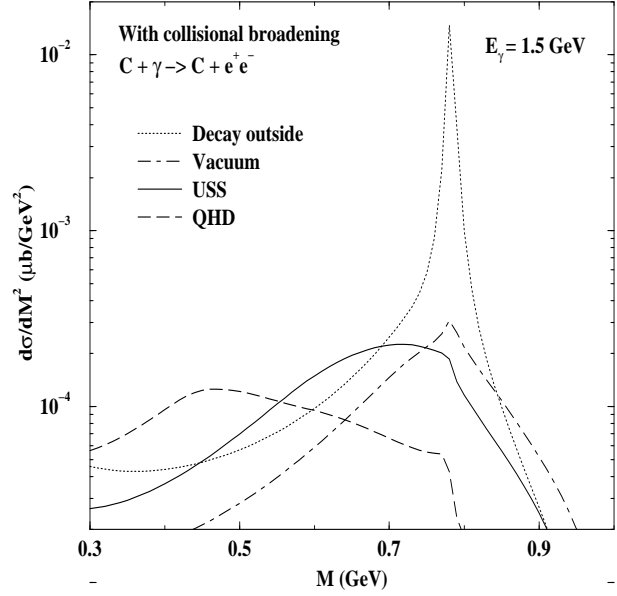,width=8cm,height=8cm}}
\caption{Same as \protect{Fig.\ref{fig5}} 
for  $E_\gamma=1.5$ GeV.
}
\label{fig8}
\end{figure}

\begin{figure}
\centerline{\psfig{figure=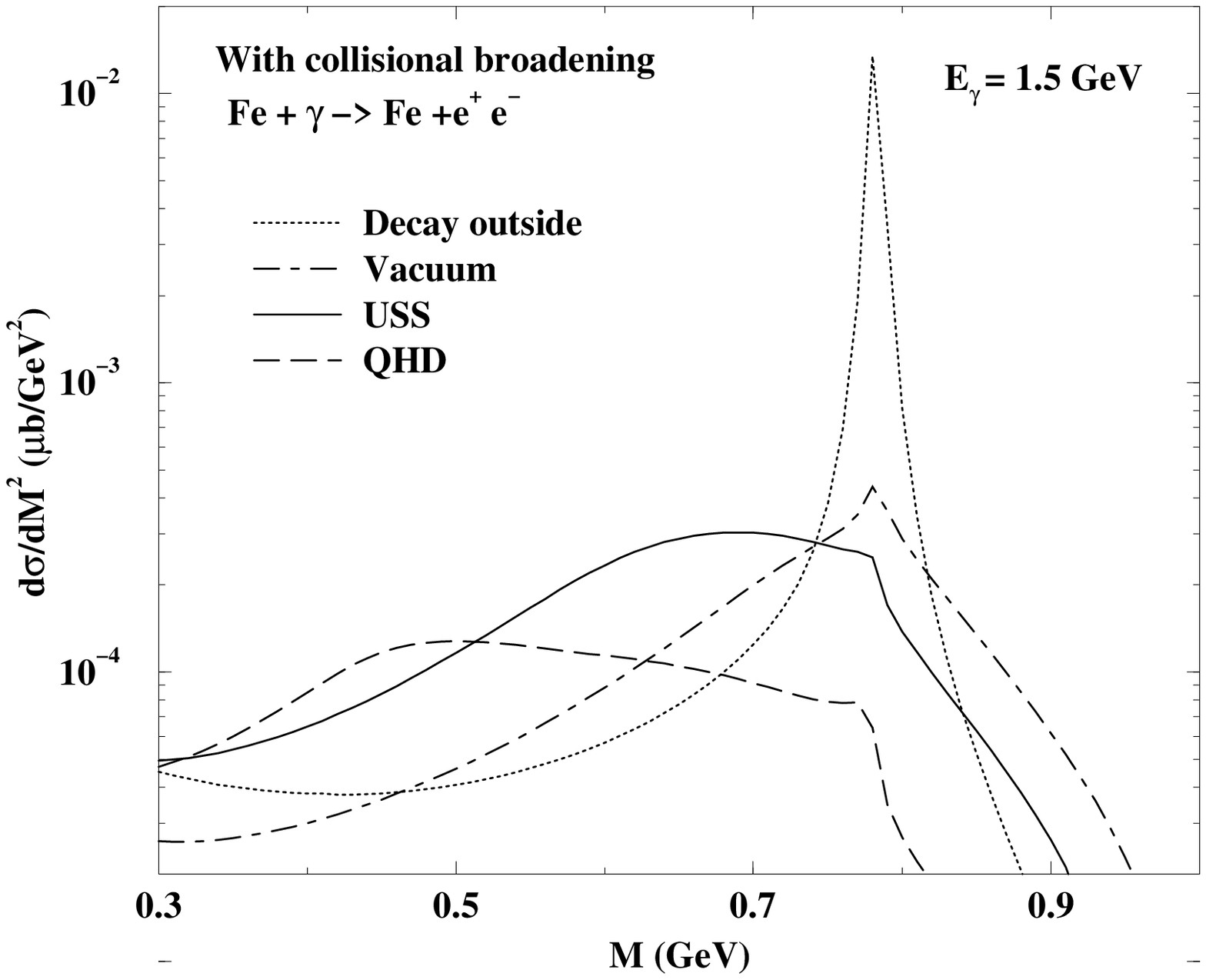,width=8cm,height=8cm}}
\caption{Same as \protect{Fig.\ref{fig6}} 
for  $E_\gamma=1.5$ GeV.
}
\label{fig9}
\end{figure}
\begin{figure}
\centerline{\psfig{figure=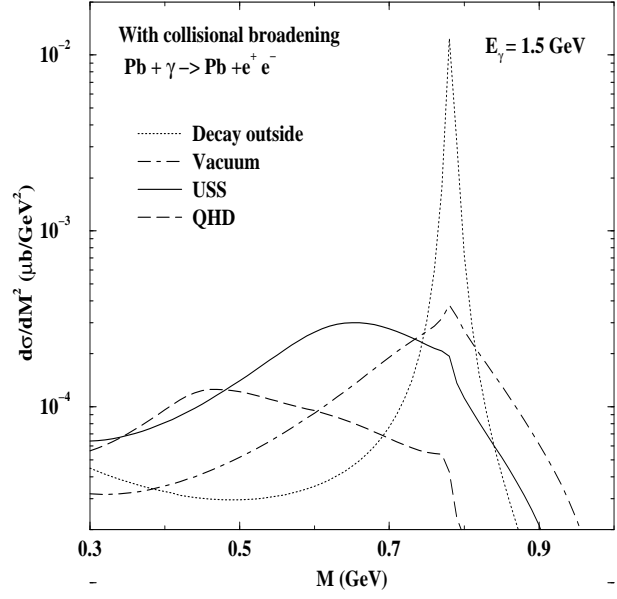,width=8cm,height=8cm}}
\caption{Same as \protect{Fig.\ref{fig7}} 
for  $E_\gamma=1.5$ GeV.
}
\label{fig10}
\end{figure}
In Figs.~\ref{fig8}, \ref{fig9} and \ref{fig10} we show the 
dilepton emission cross section as a function of the invariant mass,
$M$ for $E_\gamma=1.5$ GeV. At this energy $\rho$ and $\omega$
mesons may be created inside the nucleus with non-zero velocity.
$E_\gamma=1.5$ GeV is  below the $\phi$ production threshold.
The decay probability of the vector meson outside the nuclear
volume increases with incident photon energy.
For smaller nucleus (carbon, say) the dilepton spectra 
for the vector mesons decaying inside is marginally larger
than the spectra originating from the decays outside for 
$M<0.5$ GeV since
the vector meson with non-zero velocity has a probability 
to traverse the (small) nucleus before decaying. However, for
larger nuclei (Fe and Pb) the decay probability inside
the nucleus is more. Consequently, the dilepton spectra
from the vector mesons decaying inside dominates over the
one corresponding to the decay outside for invariant mass
region $0.3<M$(GeV)$<0.7$ for USS and QHD scenario.  

\begin{figure}
\centerline{\psfig{figure=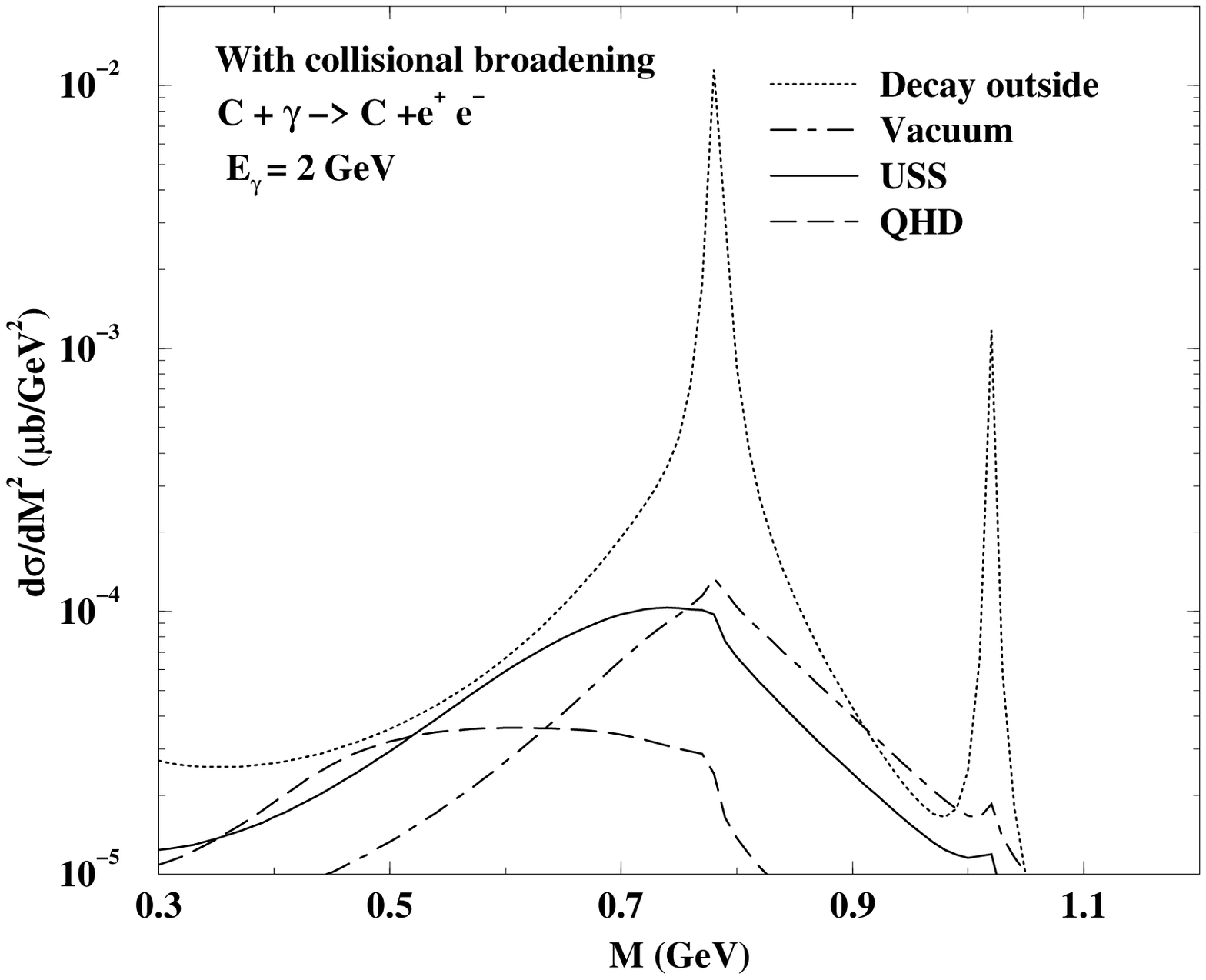,width=8cm,height=8cm}}
\caption{Same as \protect{Fig.\ref{fig5}} 
for  $E_\gamma=2$ GeV.
}
\label{fig11}
\end{figure}
\begin{figure}
\centerline{\psfig{figure=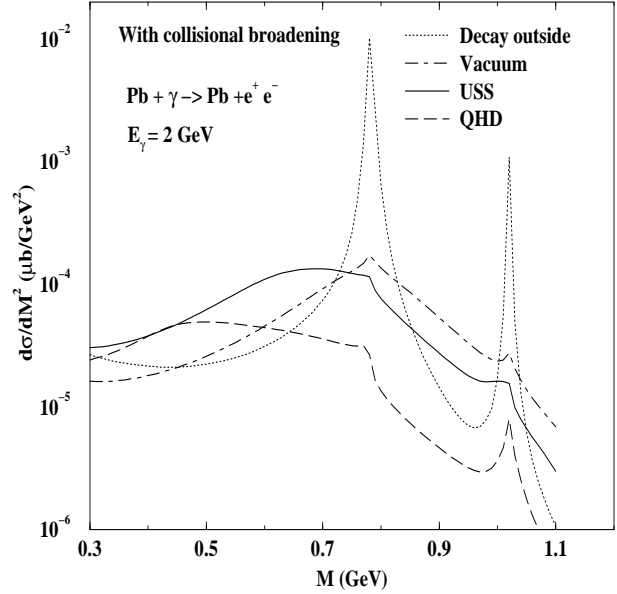,width=8cm,height=8cm}}
\caption{Same as \protect{Fig.\ref{fig7}} 
for  $E_\gamma=2$ GeV.
}
\label{fig12}
\end{figure}
The dilepton spectra for $E_\gamma=2$ GeV are shown in
Figs.~\ref{fig11} and \ref{fig12} for C and Pb respectively. 
At this energy all the three low mass vector mesons
($\rho$, $\omega$ and $\phi$) are produced with finite
velocity. The are likely to pass through the whole
nucleus and decay outside. This is reflected in the invariant 
mass distribution of dileptons from carbon target, where most
of the vector mesons decay outside the nuclear volume.
In case of Pb target, however, the lepton pairs from vector mesons
decaying inside dominates in the low invariant mass region.

\begin{figure}
\centerline{\psfig{figure=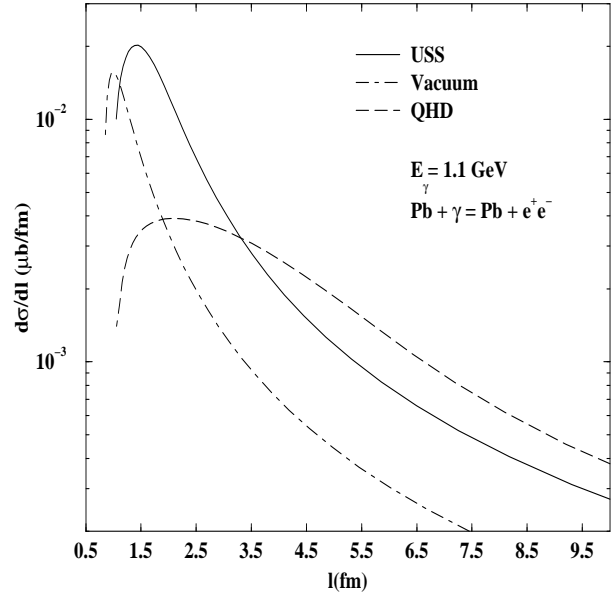,width=8cm,height=8cm}}
\caption{The dilepton spectra as a function of average path length
inside the Pb nucleus for  $E_\gamma=1.1$ GeV.
}
\label{fig13}
\end{figure}

\begin{figure}
\centerline{\psfig{figure=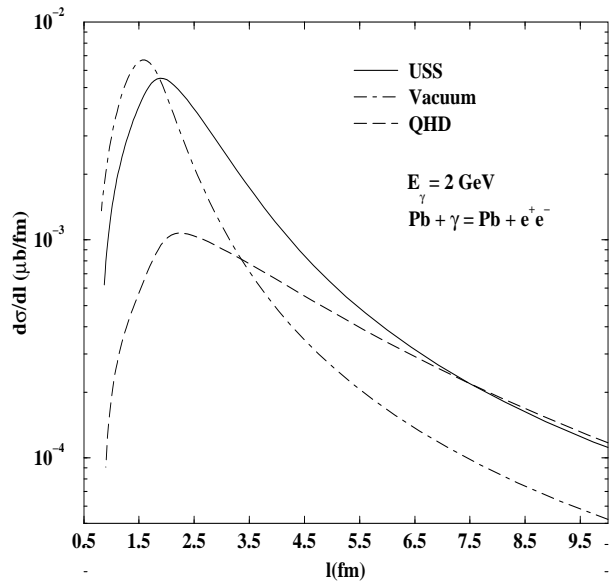,width=8cm,height=8cm}}
\caption{The dilepton spectra as a function of average path length
inside the Pb nucleus for  $E_\gamma=2$ GeV.
}
\label{fig14}
\end{figure}

In Figs.~\ref{fig13} and \ref{fig14} we plot 
the invariant mass distribution of lepton pairs as a function
of the path length traveled by the vector mesons inside the nucleus for
$E_\gamma=1.1$  and 2 GeV respectively. The path length
of the vector meson can be calculated by using the equation
$l=\frac{\mid\vec{p_3}\mid}{M}/\Gamma_{tot}$. This 
has been averaged over the Mandelstam variable $t$
(contained in $p_3$). $l$ indicates the distance
inside the nucleus probed by the vector meson. 
In view of the fact that 
at a given nuclear
density the $\rho$  is lighter in QHD than USS,
it is clear from the results shown in 
Figs.~\ref{fig13} and \ref{fig14} that lighter the mesons the more 
it penetrates the target.

%

\section{Summary and Discussions}
In this work we have studied the effects of the spectral shift
of vector mesons on the photon-nucleus interactions. It is found
that the experimental data on nuclear shadowing can be well described
within the ambit of universal scaling scenario proposed by Brown and
Rho. The spectral shift of vector mesons in Quantum Hadrodynamical model
seems to underestimate the data. On the other hand vector mesons with
their vacuum properties overestimate the data. 
However, we emphasize that 
our understanding of the shadowing phenomena vis-a-vis 
Glauber model may be improved through corrections to the 
approximations inherent in the model as well as by examining 
the model parameters critically.
The leading correction to the Glauber model due to deviation
from eikonal propagation gives rise to a correction $\sim (2P_{cm}^AR)^{-1}$
relative to the Glauber scattering amplitude~\cite{sjw}. Here $P_{cm}^A$ is the
centre of mass momentum of the nucleus and $R$ is its charge
radius. In the present case the correction is rather small
for $E_\gamma>1$ GeV. Moreover, a refinement of the Glauber model (multiple
scattering) parameters, {\it e.g.}, vector meson-nucleon scattering
amplitude, two nucleon correlation etc. might give a good agreement with
the data even with the vacuum properties of the hadrons. Hence a better
estimate of these quantities is essential for
a definitive statement regarding the role of medium effects on  
shadowing in photo-absorption processes.  
Experimental data with better statistics would certainly help 
us to resolve these uncertainties.

We have shown that the modification of the hadronic spectral
function inside the nucleus drastically change the invariant
mass distribution of lepton pairs originating from the decays
of the vector mesons. To bring out the effect of different 
densities and hence effective in-medium masses and widths, 
we have explicitly incorporated the co-ordinate dependence for
production and decay points. The dilepton spectra originating 
from the vector meson decays show strong dependence on the
incident photon energy and mass number of the target nuclei.
For this purpose we have considered incident photon energies of
1.1, 1.5 and 2 GeV on carbon, iron and lead nuclei.
It has been demonstrated that by
tuning the incident photon energy one can create the vector
meson inside the nucleus with very small velocity. This
will facilitate it to decay within the nuclear volume and leptons
being electromagnetically interacting particles should carry
the information of the in-medium properties of the vector mesons 
very efficiently.  We have also shown that heavier the nucleus 
more visible are the medium effects.  

It is to be remembered that the ``signal''
for the modification of the vector mesons properties should be
filtered out from the background from Dalitz decays and decay
of the vector mesons outside the nucleus. By fixing the beam 
energy one can minimize the later contributions. For the 
estimation of the backgrounds
from Dalitz decays one needs to know the distributions of the
hadrons {\it e.g.} $\pi$, $\eta$, $\omega$, $\Delta$'s etc
produced in $\gamma$-A collision.
The contributions from Bethe-Heitler process can be
eliminated by suitable cuts in the final spectra.

\section*{Appendix}
\setcounter{equation}{0}
\renewcommand{\theequation}{A.\arabic{equation}}

\subsection{Differential cross-section for $2 \rightarrow 3$ particle scattering}

We treat the $2 \rightarrow 3$ scattering process on the basis 
of the factorization
of the phase space integral into two processes $2 \rightarrow 2$ 
and $1 \rightarrow 2$. We 
choose the two-particle intermediate system to be $k_1+k_2$. The differential
cross-section for the process $p_1 +p_2 \rightarrow p_3 (k_1+k_2)+p_4$ is 
given by~\cite{book},
\be
\sigma_{l^+l^-} = \int\,\frac{\hl {|\cm|}^2\,d{\rm Lips}}{\cal {F}},
\label{ap1}
\ee
where ${\cal {F}}$ is the flux factor given by ${\cal {F}} = 2\,
\lambda(s,m_1^2,m_2^2)$ 
with $\lambda(x,y,z)=x^2+y^2+z^2-2(xy+yz+zx)$.
The Lorentz-invariant phase space factor is
given by
\be
\int\,d{\rm Lips} =\int\,\frac{d^3p_4}{(2\pi)^3 2E_4}\,\frac{d^3k_1}
{(2\pi)^3 2\omega_1}\,\frac{d^3k_2}{(2\pi)^3 2\omega_2}\,
(2\pi)^4\,\delta^4(p_1+p_2-p_4-k_1-k_2). 
\label{ap2}
\ee
In order to write the intermediate state explicitly in the phase space 
integral, the identity
\be
1 = \int\,ds_2\,\int\,\frac{d^3p_3}{2E_3}\,\delta^4(p_3-k_1-k_2)
\label{ap3}
\ee
with $E_3^2 = {\bf p}_3^2+s_2$ is used in Eq.~(\ref{ap2}) to obtain
\bea
\int\,d{\rm Lips}& =&\frac{1}{(2\pi)^5} \int s_2\,\left\{\int\,\frac{d^p_4}{2E_4}\,
\frac{d^3p_3}{2E_3}\,\delta^4(p_1+p_2-p_3-p_4)\right\}\nonumber\\
&&\times\,\left\{\int\,\frac{d^3k_1}{2\omega_1}\,\frac{d^3k_2}
{\omega_2}\,\delta^4(p_3-k_1-k_2)\right\}\nonumber\\
&=&\frac{1}{(2\pi)^5}\,\int\,ds_2\,R_2(s,m_4^2,s_2)\,R_2(s_2,m_{k_1}^2,m_{k_2}^2)
,
\label{ap4}
\eea
where $s = (p_1+p_2)^2$.
Let us consider the first two particle phase space $R_2(s,m_4^2,s_2)$.
Going to CMS: ${\bf p}_1+{\bf p}_2 = 0$ one obtains
\bea
R_2(s,m_4^2,s_2)& =& \frac{\pi}{2}\,\int\,\frac{p_4^2\,dp_4\,d\cos\theta_{24}}
{4E_4 E_3}\,\delta(E_1+E_2-E_4-E_3)\nonumber\\
&=& \frac{\pi}{2 \sqrt{s}}\,p_{4 {\rm cm}}\int\,d\cos{\theta_{24}} 
\nonumber\\
&=&\frac{\pi}{2 \lambda^{1/2}(s,m_1^2,m_2^2)}\,\int\,dt_1,
\label{ap5}
\eea
with $t_1 = (p_1-p_3)^2 = (p_2-p_4)^2 = m_2^2+m_4^2-2E_2 E_4+2p_{2{\rm cm}}
p_{4{\rm cm}} \cos{\theta_{24}}$.

The second two-particle phase space integral, in the rest frame of ${\bf k}_1
+{\bf k}_2$ (quantities denoted by R), can be wriiten as
\bea 
R_2(s_2,m_{k_1}^2,m{k_2}^2) &=& \int\,\frac{d^3k_1}{4\omega_1 \omega_2}\,
\delta(E_3-\omega_1-\omega_2)\nonumber\\
&=&2\pi\,\frac{\lambda^{1/2}(s_2,m_{k_1}^2,m_{k_2}^2)}{8s_2}\int\,
d\cos{\theta_{k_1p_1}^R}
\label{ap6}
\eea
We define the invariant variable $t_2$ as follows:
\bea
t_2& =& (p_1-k_1)^2 = m_1^2+m_{k_1}^2-2E_1^R \omega_1^R+2p_1^R k_1^R 
\cos{\theta_{k_1 p_1}^R}
\label{ap7}
\eea
so that $dt_2 = \lambda^{1/2}(s_2,m_1^2,t_1)\,\lambda^{1/2}(s_2,m_{k_1}^2,
m_{k_2}^2)\,d\cos{\theta_{k_1p_1}^R}/2s_2$ and we obtain from Eq.~(\ref{ap6})
\be
R_2(s_2,m_{k_1}^2,m_{k_2}^2) = \frac{\pi}{2\lambda^{1/2}(s_2,m_1^2,t_1)}\,
\int\,dt_2
\label{ap8}
\ee
Finally, combining Eqs.~(\ref{ap1}), (\ref{ap4}), (\ref{ap5}) and (\ref{ap8})
the dilepton production cross-section is obtained as
\be
\frac{d\sigma_{l^+ l^-}}{dM^2\,dt_1\,dt_2}=
\frac{|\hl{\cm}(M^2,t_1,t_2)|^2}{256\pi^3\,\lambda(s,0,m_N^2)
\,\lambda^{1/2}(M^2,0,t_1)}
\ee
The limits of $t_1$ and $t_2$ are given by
\bea
t_1|^{max}_{min}&=&2m_N^2-\frac{1}{2s}\left[(s+m_N^2)(s-M^2+m_N^2)\mp
\lambda^{1/2}(s,m_N^2,0)\lambda^{1/2}(s,m_N^2,M^2)\right]\\
t_2|^{max}_{min}&=&-\frac{1}{2}(M^2-t_1)\pm
\left[\lambda^{1/2}(M^2,0,0)\lambda^{1/2}(M^2,t_1,0)\right]
\eea

\subsection*{B. Invariant amplitudes}

The invariant amplitudes 
for the Feynman diagrams shown in 
Fig.~(\ref{fig2})(a-e) 
are given by
\be
\cm_a=\frac{-ieF_\pi\ggo\ggr\gpnn\gwrp\,
\eps^{\mu\nu\alpha\beta}p_{1\mu}\epsilon^\gamma_\nu(p_1)(k_1+k_2)_{\alpha}\,
\bu_l(k_2)\gamma_\beta v_l(k_1)\,\bu_N(p_4)\gamma_5 u_N(p_2)}
{M^2\mpi [t_1-\mpi^2][\mo^2-i\mo\Go][M^2-\mr^2+i\mr\Gr]}
\ee
\be
\cm_b=\frac{-ieF_\pi\ggr\ggo\gpnn\gwrp\,
\eps^{\mu\nu\alpha\beta}p_{1\mu}\epsilon^\gamma_\nu(p_1)(k_1+k_2)_{\alpha}\,
\bu_l(k_2)\gamma_\beta v_l(k_1)\,\bu_N(p_4)\gamma_5 u_N(p_2)}
{M^2\mpi [t_1-\mpi^2][\mr^2-i\mr\Gr][M^2-\mo^2+i\mo\Go]}
\ee
\be
\cm_c=\frac{-ieF_\pi\ggp\ggr\gpnn\gprp\,
\eps^{\mu\nu\alpha\beta}p_{1\mu}\epsilon^\gamma_\nu(p_1)(k_1+k_2)_{\alpha}\,
\bu_l(k_2)\gamma_\beta v_l(k_1)\,\bu_N(p_4)\gamma_5 u_N(p_2)}
{M^2\mpi [t_1-\mpi^2][\mph^2-i\mph\Gp][M^2-\mr^2+i\mr\Gr]}
\ee
\be
\cm_d=\frac{-ieF_\pi\ggr\ggp\gpnn\gprp\,
\eps^{\mu\nu\alpha\beta}p_{1\mu}\epsilon^\gamma_\nu(p_1)(k_1+k_2)_{\alpha}\,
\bu_l(k_2)\gamma_\beta v_l(k_1)\,\bu_N(p_4)\gamma_5 u_N(p_2)}
{M^2\mpi [t_1-\mpi^2][\mr^2-i\mr\Gr][M^2-\mph^2+i\mph\Gp]}
\ee
\be
\cm_e=\frac{-2eF_\sigma\ggr^2\gsrr\gsnn\,
\epsilon^\gamma_\mu(p_1)[p_1\cdot (k_1+k_2)g^{\mu\beta}-p_1^\beta (k_1+k_2)^\mu]
\bu_l(k_2)\gamma_\beta v_l(k_1)\,\bu_N(p_4)u_N(p_2)}
{M^2\mpi [t_1-\ms^2+i\ms\Gs][\mr^2-i\mr\Gr][M^2-\mr^2+i\mr\Gr]}
\ee

The squared invariant amplitude is then
\be
\hl{|\cm|}^2=\hl{|\cm_a+\cm_b+\cm_c+\cm_d|}^2+\hl{|\cm_e|}^2
\ee
where
\be
\hl{|\cm_a|}^2=\frac{F_\pi^2\,A_\pi\,\ggo^2\ggr^2\gpnn^2\gwrp^2}
{[\mo^4+\mo^2\Go^2][(M^2-\mr^2)^2+\mr^2\Gr^2]}
\ee
\be
\hl{|\cm_b|}^2=\frac{F_\pi^2\,A_\pi\,\ggr^2\ggo^2\gpnn^2\gwrp^2}
{[\mr^4+\mr^2\Gr^2][(M^2-\mo^2)^2+\mo^2\Go^2]}
\ee
\be
\hl{|\cm_c|}^2=\frac{F_\pi^2\,A_\pi\,\ggp^2\ggr^2\gpnn^2\gprp^2}
{[\mph^4+\mph^2\Gp^2][(M^2-\mr^2)^2+\mr^2\Gr^2]}
\ee
\be
\hl{|\cm_d|}^2=\frac{F_\pi^2\,A_\pi\,\ggr^2\ggp^2\gpnn^2\gprp^2}
{[\mr^4+\mr^2\Gr^2][(M^2-\mph^2)^2+\mph^2\Gp^2]}
\ee
\bea
Re\,\hl{\cm_a^\ast\cm_b}&=&\frac{F_\pi^2\,A_\pi\,\ggo^2\ggr^2\gpnn^2\gwrp^2}
{[\mo^4+\mo^2\Go^2][(M^2-\mr^2)^2+\mr^2\Gr^2]
[\mr^4+\mr^2\Gr^2][(M^2-\mo^2)^2+\mo^2\Go^2]}\nonumber\\
&\times&\left[\frac{}{}\left(\mo^2\mr^2+\mo\mr\Go\Gr\right)
\left\{(M^2-\mr^2)(M^2-\mo^2)+\mo\mr\Go\Gr\right\}\right.\nonumber\\
&&\left.-\left(\mr^2\mo\Go-\mo^2\mr\Gr\right)
\left\{(M^2-\mo^2)\mr\Gr-(M^2-\mr^2)\mo\Go\right\}\frac{}{}\right]
\eea
\bea
Re\,\hl{\cm_a^\ast\cm_c}&=&\frac{F_\pi^2\,A_\pi\,\ggo\ggp\ggr^2\gpnn^2
\gwrp\gprp}
{[\mo^4+\mo^2\Go^2][(M^2-\mr^2)^2+\mr^2\Gr^2]^2
[\mph^4+\mph^2\Gp^2]}\nonumber\\
&\times&\left[\frac{}{}\left(\mo^2\mph^2+\mo\mph\Go\Gp\right)
\left\{(M^2-\mr^2)^2+\mr^2\Gr^2\right\}\right]
\eea
\bea
Re\,\hl{\cm_a^\ast\cm_d}&=&\frac{F_\pi^2\,A_\pi\,\ggo\ggp\ggr^2
\gpnn^2\gwrp\gprp}
{[\mo^4+\mo^2\Go^2][(M^2-\mr^2)^2+\mr^2\Gr^2]
[\mr^4+\mr^2\Gr^2][(M^2-\mph^2)^2+\mph^2\Gp^2]}\nonumber\\
&\times&\left[\frac{}{}\left(\mo^2\mr^2+\mo\mr\Go\Gr\right)
\left\{(M^2-\mr^2)(M^2-\mph^2)+\mr\mph\Gr\Gp\right\}\right.\nonumber\\
&&\left.-\left(\mr^2\mo\Go-\mo^2\mr\Gr\right)
\left\{(M^2-\mph^2)\mr\Gr-(M^2-\mr^2)\mph\Gp\right\}\frac{}{}\right]
\eea
\bea
Re\,\hl{\cm_b^\ast\cm_c}&=&\frac{F_\pi^2\,A_\pi\,\ggo\ggp\ggr^2
\gpnn^2\gwrp\gprp}
{[\mr^4+\mr^2\Gr^2][(M^2-\mr^2)^2+\mr^2\Gr^2]
[\mph^4+\mph^2\Gp^2][(M^2-\mo^2)^2+\mo^2\Go^2]}\nonumber\\
&\times&\left[\frac{}{}\left(\mr^2\mph^2+\mr\mph\Gr\Gp\right)
\left\{(M^2-\mo^2)(M^2-\mr^2)+\mo\mr\Go\Gr\right\}\right.\nonumber\\
&&\left.-\left(\mr^2\mph\Gp-\mph^2\mr\Gr\right)
\left\{(M^2-\mo^2)\mr\Gr-(M^2-\mr^2)\mo\Go\right\}\frac{}{}\right]
\eea
\bea
Re\,\hl{\cm_b^\ast\cm_d}&=&\frac{F_\pi^2\,A_\pi\,\ggo\ggp\ggr^2
\gpnn^2\gwrp\gprp}
{[\mr^4+\mr^2\Gr^2][(M^2-\mo^2)^2+\mo^2\Go^2]
[(M^2-\mph^2)^2+\mph^2\Gp^2]}\nonumber\\
&\times&\left[\frac{}{}(M^2-\mo^2)(M^2-\mph^2)+\mo\mph\Go\Gp\right]
\eea
\bea
Re\,\hl{\cm_c^\ast\cm_d}&=&\frac{F_\pi^2\,A_\pi\,\ggp^2\ggr^2\gpnn^2\gprp^2}
{[\mr^4+\mr^2\Gr^2][(M^2-\mr^2)^2+\mr^2\Gr^2]
[\mph^4+\mph^2\Gp^2][(M^2-\mph^2)^2+\mph^2\Gp^2]}\nonumber\\
&\times&\left[\frac{}{}\left(\mr^2\mph^2+\mr\mph\Gr\Gp\right)
\left\{(M^2-\mr^2)(M^2-\mph^2)+\mr\mph\Gr\Gp\right\}\right.\nonumber\\
&&\left.-\left(\mr^2\mph\Gp-\mph^2\mr\Gr\right)
\left\{(M^2-\mph^2)\mr\Gr-(M^2-\mr^2)\mph\Gp\right\}\frac{}{}\right]
\eea
\be
\hl{|\cm_e|}^2=\frac{F_\sigma^2\,A_\sigma\,\ggr^4\gsnn^2\gsrr^2}
{[\mr^4+\mr^2\Gr^2][(M^2-\mr^2)^2+\mr^2\Gr^2]}
\ee
with
\[
A_\pi=-\frac{2\pi\alpha}{M^2}\frac{\left[t_2^2+(t_2-t_1+M^2)^2\right]t_1}
{\mpi^2[t_1-\mpi^2]^2}~~,~~
A_\sigma=\frac{8\pi\alpha}{M^2}\frac{\left[t_2^2+(t_2-t_1+M^2)^2\right]
(4m_N^2-t_1)}{\mpi^2[(t_1-\ms^2)^2+\ms^2\Gs^2]^2}~,
\]
and
\[
F_\pi=\left[\frac{\Lambda_\pi^2-\mpi^2}{\Lambda_\pi^2-t_1}\right]
\left[\frac{\mr^2-\mpi^2}{\mr^2-t_1}\right]~~,~~~
F_\sigma=\left[\frac{\Lambda_\sigma^2-\ms^2}{\Lambda_\sigma^2-t_1}\right]
\left[\frac{\Lambda^2_{\sigma\rho\rho}-\mr^2}{\Lambda^2_{\sigma\rho\rho}-t_1}
\right]~.
\]

\end{document}